\documentclass[aps,prb,preprint,groupedaddress]{revtex4-1}
\usepackage{graphicx}
\usepackage{amsmath}
\usepackage[dvipdfm]{}
\def\tc{$T_c$} \def\tl{$1/T_1$} \def\kb{$k_BT_c$} 

\begin{document}
\title{
Spin-singlet superconductivity with a full gap in \\ locally non-centrosymmetric SrPtAs}


\author{K. Matano$^{1}$, K. Arima$^1$, S. Maeda$^1$, Y. Nishikubo$^1$, K. Kudo$^{1}$, M. Nohara$^{1}$, and Guo-qing Zheng$^{1,2}$}
\affiliation{
$^1$Department of Physics, and Research Center of New Functional Materials for Energy Production,\\
 Storage and Transport, Okayama University, Okayama 700-8530, Japan\\
$^2$Institute of Physics and Beijing National Laboratory for Condensed Matter Physics, Chinese Academy of Sciences, Beijing 100190, China
}
\date{\today}

\begin{abstract}
We report $^{195}$Pt-NMR and $^{75}$As-NQR measurements for the locally non-centrosymmetric superconductor SrPtAs
where the As-Pt layer breaks inversion symmetry while globally the compound is centrosymmetric.
The nuclear spin lattice relaxation rate $1/T_1$ shows a well-defined coherence peak below $T_c$ and decreases exponentially at low temperatures.
The spin susceptibility measured by the Knight shift also decreases below $T_c$ down to $T<T_c/6$. 
These data together with the penetration depth obtained from the NMR spectra can be consistently explained by assuming a spin-singlet superconducting state with a full gap.
Our results suggest that the spin-orbit coupling due to the local inversion-breaking is not large enough to bring about an exotic superconducting state, 
or the inter-layer hopping interaction is larger than the spin-orbit coupling.
\end{abstract}

\pacs{74.25.N−, 71.70.Ej, 74.25.Jb, 76.60.Cq}

\maketitle
In non-centrosymmetric superconductors, an antisymmetric spin-orbit coupling (ASOC) interaction is induced.
As a result, a parity mixed superconducting state is allowed and the mixing extent is determined by the strength of an ASOC.
\cite{PhysRevLett.87.037004,PhysRevLett.92.097001,1367-2630-6-1-115} 
Indeed, some unconventional superconductors such as spin-triplet, nodal-gap superconductivity have been reported.
For example, a nodal-gap and spin-triplet superconducting state has been reported in Li$_2$Pt$_3$B.\cite{Nishiyama_PhysRevLett.98.047002,PhysRevLett.97.017006}
However, Li$_2$Pt$_3$B is the only example that shows unconventional superconductivity and other superconductors containing 
heavy elements are conventional.\cite{Nishiyama_PhysRevB.71.220505,
Tahara_PhysRevB.80.060503,Mo3Al2C_transport_Bauer,PhysRevB.85.052501,matano_JPSJ.82.084711}
The difference was explained by the peculiar crystal structure distortion as to increase the extent of inversion-symmetry breaking in  Li$_2$Pt$_3$B.
\cite{Harada_PhysRevB.86.220502}


Recently, ``locally'' non-centrosymmetric systems have also attracted attention.\cite{PhysRevB.84.184533,JPSJ.81.034702}
In these systems, the whole structure remains centrosymmetric, but inversion symmetry is broken locally in some parts within the unit cell.
In such systems, a possible exotic superconducting state is suggested.\cite{PhysRevB.84.184533,JPSJ.81.034702}
However, these proposals have not been tested because of lack of samples.
SrPtAs ($T_c$ $\sim$ 2.4 K) is one of such candidates.\cite{SrPtAs} 

SrPtAs has a honeycomb layered structure and consists of Sr and Pt-As layers.
SrPtAs has an inversion symmetry in the whole unit cell, but the Pt-As layer lacks an inversion center.
According to the theoretical calculation, a sizable ASOC is expected.\cite{PhysRevB.85.220505}
When the interlayer hopping is smaller than the ASOC, an exotic state, such as a chiral $d$-wave or $f$-wave state can be expected.
\cite{PhysRevB.86.100507,2013arXiv1312.3071W,PhysRevB.85.220505}
In this paper, we report the $^{195}$Pt nuclear magnetic resonance (NMR) and 
$^{75}$As nuclear quadrupole resonance (NQR) measurements in the superconducting and the normal states. 
The spin-lattice relaxation rate, \tl, shows a clear coherence peak just below \tc\ and decays exponentially with decreasing $T$,
indicating a fully opened superconducting gap on the whole Fermi surface.
The Knight shift decreases below \tc, indicating a spin-singlet pairing.
Our results suggest that the inter-layer hopping interaction is larger than the spin-orbit coupling due to the local inversion-breaking
or the spin-orbit coupling is 
not large enough to bring about an exotic superconducting state, as opposed to the theoretical proposals.

The  polycrystalline sample of SrPtAs was synthesized by a solid-state reaction. 
The PtAs$_2$ precursor was first synthesized by heating Pt powder and As grains at 700 $^\circ$C in an evacuated quartz tube. 
Then, Sr, Pt, and PtAs$_2$ powders of stoichiometric amounts were mixed and ground. 
The resulting powder was placed in an alumina crucible and sealed in an evacuated quartz tube. 
The ampule was heated at 700 $^\circ$C for 3 h and then at 1000 $^\circ$C for 24 h. 
After furnace cooling, the sample was ground, pelletized, wrapped with Ta foil, and heated at 950 $^\circ$C for 2 h in an evacuated quartz tube.\cite{SrPtAs}
The pellet was crushed into powders for NMR/NQR measurements.
The $T_c$ at zero and a finite magnetic field $H$ was determined 
by measuring the ac susceptibility using the {\it in situ} NMR/NQR coil.
The $T_c$ is 2.40 K at zero field and 1.43 K at 0.0842 T.
A standard phase-coherent pulsed NMR spectrometer was used to collect data.
NMR/NQR measurements were performed using the spin echo method.
The nuclear spin-lattice relaxation rate was measured by using a single saturation pulse.
Measurements below 1.4 K were carried out in a $^3$He-$^4$He dilution refrigerator.
 \begin{figure}[htbp]
 \includegraphics[clip,width=65mm]{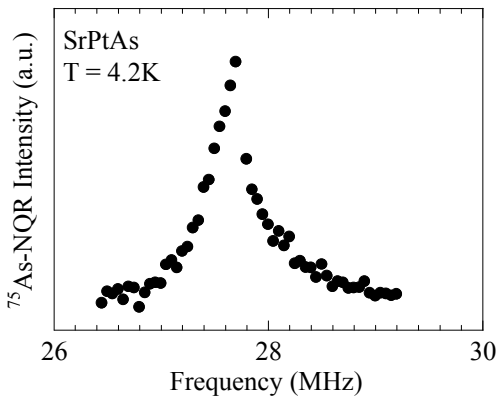}
 \caption{\label{f1}$^{75}$As-NQR spectrum of SrPtAs measured at $T$ = 4.2 K.}
 \end{figure}

Figure \ref{f1} shows the $^{75}$As ($I = 3/2$) NQR spectrum at $T$ = 4.2 K. 
The sharp peak centered at $f$ = 27.5MHz.
The full width at the half maximum (FWHM) of the NQR spectrum is 0.54 MHz.
\begin{figure}[htbp]
 \includegraphics[clip,width=70mm]{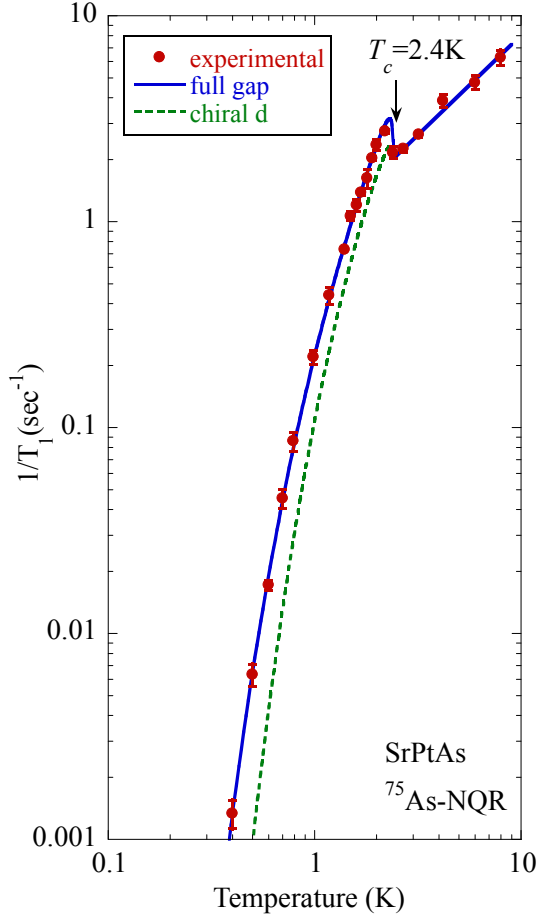}
 \caption{\label{f2}
 (color online) Temperature dependence of the $^{75}$As spin-lattice relaxation rate, 
 \tl\, measured by NQR.
The straight line above \tc\ represents the $T_1T$ = const relation. 
The solid curve below \tc\ is a calculation assuming the BCS function. The dotted curve is a calculation for a chiral $d$-wave state (see text for detail).}
 \end{figure}
 
 \begin{figure}[htbp]

 \includegraphics[clip,width=60mm]{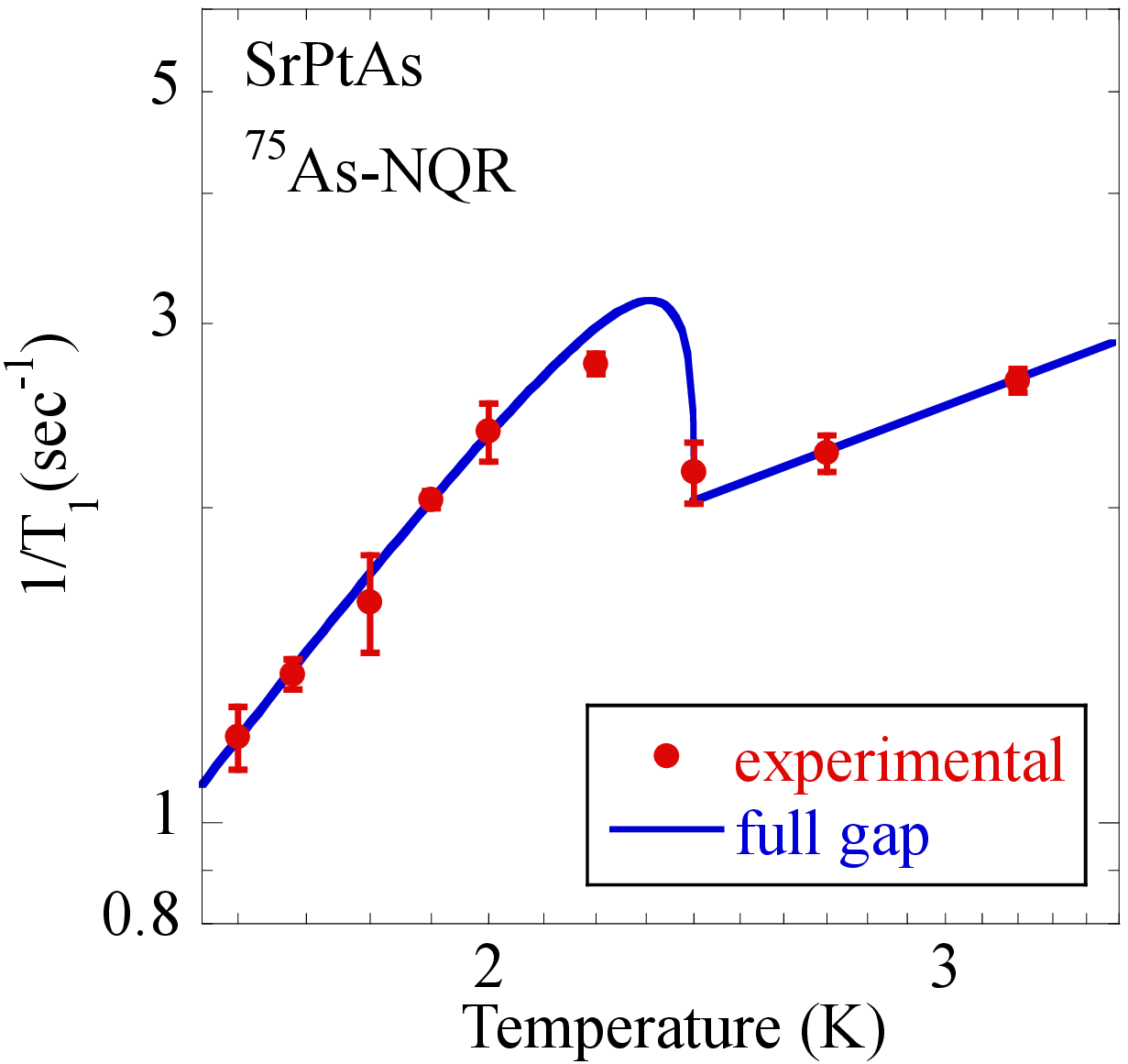}
 \caption{\label{f2-2}
 (color online) The enlarged part of \tl\ around \tc .
The straight line above \tc\ represents the $T_1T$ = const relation. 
The solid curve below \tc\ is a calculation assuming the BCS function. }
 \end{figure}
Figure \ref{f2} shows the temperature dependence of \tl\ measured at the peak of the NQR spectrum at zero magnetic field.
The nuclear magnetization curve was fitted by a single exponential function: 
\begin{eqnarray}
\frac{M_0-M(t)}{M_0} = \exp(-\frac{3t}{T_1}), 
\end{eqnarray}
where $M_0$ and $M(t)$ are the nuclear magnetization in the thermal equilibrium and at a time $t$ after the saturating pulse, respectively.
Figure \ref{f2-2} shows the enlarged part of \tl\ around \tc.
As seen in the figure, \tl\ varies in proportion to the temperature ($T$) above \tc, as expected for conventional metals, indicating no electron-electron interaction.
Below \tc, \tl\ shows a coherence peak (Hebel-Slichter peak).
The $1/T_{1S}$ in the superconducting state is expressed as
\begin{eqnarray}
\frac{T_{1N}}{T_{1S}}=\frac{2}{k_BT} 
\iint\left(1+\frac{\Delta^2}{EE'}\right)N_S(E)N_S(E')  \nonumber \\
\times f(E)\left[1-f(E')\right]\delta(E-E')dEdE',
\end{eqnarray}
where $1/T_{1N}$ is the relaxation rate in the normal state, $N_S(E)$ is the superconducting density of
states (DOS), $f(E)$ is the Fermi distribution function and $C =1+ \frac{\Delta^2}{EE'}$ is the coherence factor. 
To perform the calculation of eq. (2), we follow Hebel to convolute $N_S(E)$ with a broadening function $B(E)$,\cite{Hebel}
which is approximated with a rectangular function centered at $E$ with a height of 1/2$\delta$.
The solid curve below \tc\ shown in Fig. 2  is a calculation with $2\Delta = 3.85$\kb, $r\equiv\Delta(0)/\delta=4$.
It fits the experimental data reasonably well. The parameter $2\Delta$ is close to the BCS value of 3.5\kb.
This result indicates an isotropic superconducting gap in this compound.
 
Theoretically, a chiral $d$-wave superconducting state was proposed.
\cite{PhysRevB.86.100507,PhysRevB.87.180503}
In the case of $d$-wave superconductor, the coherence factor is much suppressed because of the sign change of the order parameter. 
In principle, a tiny coherence peak could be present for a $d$-wave state, but practically it is usually smeared out as seen in high-\tc\ cuprates.
\cite{PhysRevLett.100.217002,Asayama1991281}
The dotted curve of Fig. \ref{f2} and Fig. \ref{f2-2} is a calculation assuming the chiral $d$-wave with same parameter used in the $s$-wave fitting, $2\Delta = 3.85$\kb, $r=4$.
In our experiment, the coherence peak was clearly observed, which is hard to be explained by the chiral $d$-wave superconducting state.
The other report proposed a $f$-wave superconducting state.\cite{2013arXiv1312.3071W}
In such case, \tl\ should show a power law $T$ dependence due to nodes in the gap function.
So a $f$-wave superconducting state is not supported by our data. 
 \begin{figure}[htbp]
 \includegraphics[clip,width=70mm]{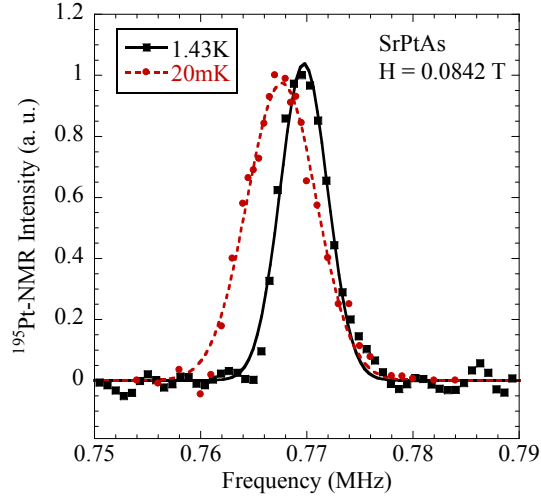}
 \caption{\label{f3}(color online) $^{195}$Pt-NMR spectrum measured at a magnetic field of $H$ = 0.0842 T above and below $T_c(H)$, respectively.}
 \end{figure}

 \begin{figure}[htbp]
 \includegraphics[clip,width=70mm]{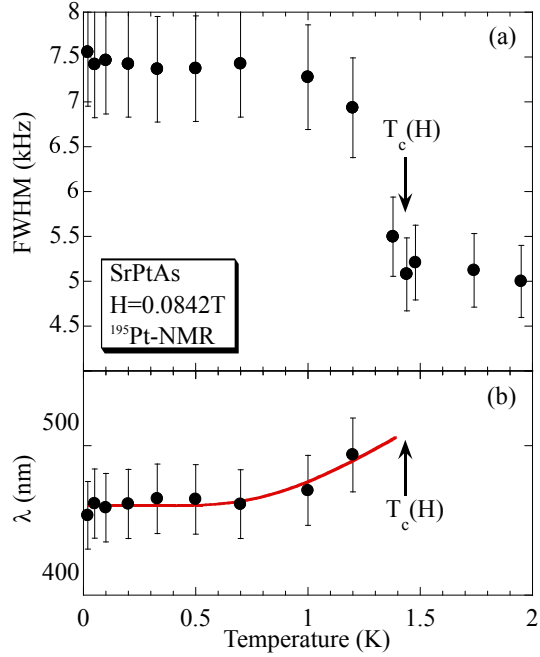}
 \caption{\label{f5}
(color online) (a) The temperature dependence of  the full width at the half maximum ($FW\!H\!M$) for the $^{195}$Pt-NMR spectrum. 
(b) The temperature dependence of the penetration depth $\lambda$. The solid curve  is a calculation by assuming a conventional $s$-wave superconducting state (see text for detail).}
 \end{figure} 

 \begin{figure}[htbp]
 \includegraphics[clip,width=70mm]{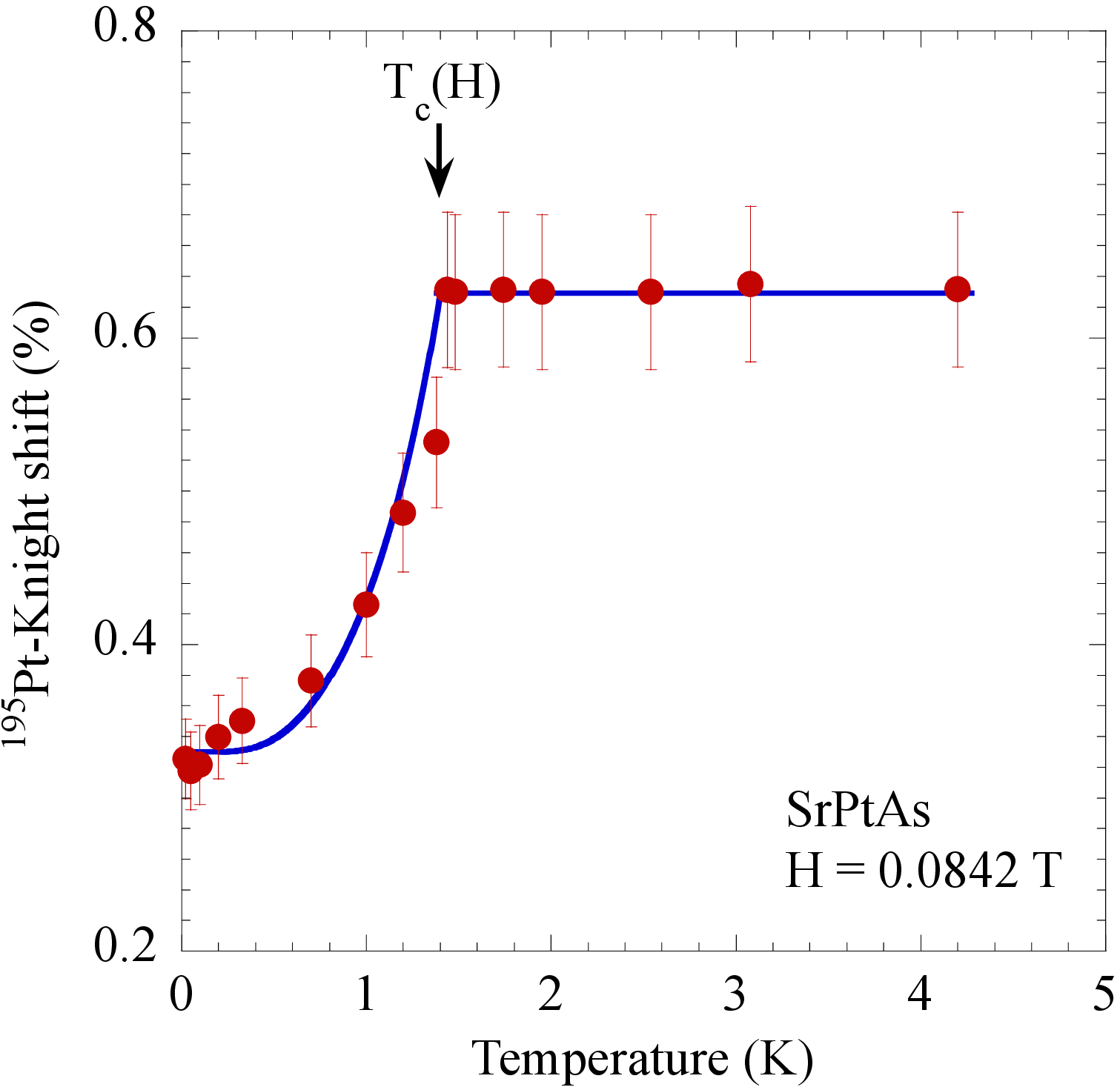}
 \caption{\label{f4}
(color online) The temperature dependence of the Knight shift for SrPtAs measured in $H = 0.0842$T.
The solid curve below \tc\ is a calculation assuming a spin-singlet pairing(see text for detail).}
 \end{figure}
Figure \ref{f3} shows the $^{195}$Pt-NMR spectra for temperature above and below $T_c(H) = 1.43$ K, respectively.
The spectra can be fitted by a single Gaussian function and the FWHM of the NMR line at 1.43 K is 5.1 kHz.
The very sharp transition indicates the high quality of the sample.
The temperature dependence of the FWHM is shown in Fig. \ref{f5}(a).
Above \tc, the FWHM is temperature independent within the error bar.
However, the FWHM is enlarged below $T{_c}(H)$.
Below \tc, a vortex state produces a distribution of the magnetic field in a superconductor.
As a result, an NMR spectrum is broadened.\cite{PhysRevLett.88.077003} The difference of the FWHM  between temperatures below and above \tc\
is related to the penetration depth as the following,\cite{PhysRevB.37.2349}
\begin{eqnarray}
{\scriptstyle \sqrt{FW\!H\!M(T=0)^2-FW\!H\!M(T\geq T_c)^2 }}  = 0.0609\gamma_n\frac{\phi_0}{\lambda^2}
\end{eqnarray}
Here, $\gamma_n$ is the gyromagnetic ratio for a nucleus,  
$\phi_0 = 2h/e = 2.07 \times 10^{-7}$Oe/cm$^2$ is the quantized magnetic flux,
and $\lambda$ is London penetration depth. 
For a fully gapped superconductor, the temperature dependence of $\lambda$ for $T/T_c<0.5$ is described as following,\cite{lambda}
\begin{eqnarray}
\lambda(T) = \lambda(0)\left[1+\sqrt[]{\frac{\pi\Delta}{2k_BT}}\exp\left(-\frac{\Delta}{k_BT}\right) \right]
\end{eqnarray}
The solid line below \tc\ in Fig. 5(b) is the calculation by assuming a fully gapped superconducting state with $\Delta$($T=0$) obtained from the $T_1$ result 
 and $\lambda(0)$ is calculated to be 460 nm.
This value is close to a previous $\mu$SR report of $\lambda$(0) = 339 nm.\cite{PhysRevB.87.180503} 
In a superconductor with a line-node gap, the penetration depth is proportional to $T$ at low temperatures.
Our result shows that the penetration depth is $T$-independent below 0.5 \tc, which also excludes the possibility of $f$-wave state.

Figure \ref{f4} shows the Knight shift, $K$, as a function of temperature.
The Knight shift was calculated by the nuclear gyromagnetic ratio  $\gamma_N$ = 9.094 MHz/T for $^{195}$Pt.
Above \tc, the shift is $T$ independent, while it decreases below \tc.
Generally, the Knight shift is expressed as,
\begin{eqnarray}
K &&= K_{orb}+K_{s} \\
K_s &&= A_{hf}\chi_s\\
\chi_s &&= -4\mu^2_B\int N_S(E)\frac{\partial f(E)}{\partial E}dE,
\end{eqnarray}
where $K_{orb}$ is the contribution due to orbital susceptibility which is $T$-independent, 
$A_{hf}$ is the hyperfine coupling constant and $\chi_s$ is the spin susceptibility.
In the present case, $K_{orb}$ is unknown. 
The solid curve below \tc\ in Fig. \ref{f4} is a calculation by assuming a spin-singlet state with the same gap parameter obtained from \tl\ fitting, 
and attributing the shift at the lowest temperature to be $K_{orb}$.
As can be seen in the figure, the result can also be well fitted by an $s$-wave state.

In conclusion, we have performed the $^{195}$Pt-NMR and $^{75}$As-NQR measurements on the locally non-centrosymmetric superconductor SrPtAs.
We find that the spin-lattice relaxation rate \tl\ shows a coherence peak just below \tc\ 
and decreases exponentially at lower temperatures.
The spin susceptibility measured by the Knight shift decreases below \tc.
These data together with the penetration depth obtained from the NMR data indicate that the Cooper pairs are in a singlet state with an isotropic gap. 
The conventional superconducting state suggests that the inter-layer hopping is large or the ASOC is weak in SrPtAs as opposed to previous theoretical proposals.

$Note$ $added:$ After completing this work, we became aware of a \tl\ measurement using NQR by Br${\rm \ddot{u}}$ckner $et$ $al$.\cite{2013arXiv1312.6166B}
However, in their sample the coherence peak is much suppressed compared to ours. 
The \tc\ of Br${\rm \ddot{u}}$ckner $et$ $al$. is lower than ours and the FWHM of NQR spectrum is broader than our sample,\cite{SrPtAs}
which may explain the suppressed coherence peak.
\begin{acknowledgments}
We thank S. Kawasaki and Y. Minami for help in the experiments and analysis.
This work was partially supported by the ``Topological Quantum Phenomena" 
Grant-in Aid for Scientific Research on innovative Areas from MEXT of Japan (Grant No. 22103004).
\end{acknowledgments}

%

\end{document}